\documentclass[onecollarge,natbib]{svjour2}
\bibpunct{[}{]}{;}{n}{}{,} 
\smartqed  
\usepackage{graphicx}
\usepackage{bm}
\usepackage{amsmath}  
\usepackage{amssymb}
\usepackage{multirow,booktabs}
\usepackage{indentfirst}

\begin{document}

\title{Benchmark Results for Few-Body Hypernuclei
}


\author{F. Ferrari Ruffino  \and
        N. Barnea           \and
        S. Deflorian        \and
        W. Leidemann        \and
        D. Lonardoni        \and
        G. Orlandini        \and
        F. Pederiva 
}


\institute{F. Ferrari Ruffino \at
               Dipartimento di Fisica, Universit\`a di Trento \& INFN, TIFPA, I-38123 Trento, Italy \\
              \email{f.ferrariruffino@unitn.it}           
           \and
           N. Barnea \at
               Racah Institute of Physics, The Hebrew University, 91904, Jerusalem, Israel \\
           \and
           S. Deflorian, W. Leidemann, G. Orlandini, F. Pederiva \at
               Dipartimento di Fisica, Universit\`a di Trento \& INFN, TIFPA, I-38123 Trento, Italy \\
           \and
           D. Lonardoni \at
               National Superconducting Cyclotron Laboratory, Michigan State University, East Lansing, Michigan 48824, USA \& 
               Theoretical Division, Los Alamos National Laboratory, Los Alamos, New Mexico 87545, USA \\  %
}

\date{Received: date / Accepted: date}

\maketitle

\begin{abstract}
The Non-Symmetrized Hyperspherical Harmonics method (NSHH) is introduced in the hypernuclear sector and benchmarked 
with three different ab-initio methods, namely the Auxiliary Field Diffusion Monte Carlo method, the Faddeev-Yakubovsky 
approach and the Gaussian Expansion Method. Binding energies and hyperon separation energies of three- to five-body 
hypernuclei are calculated by employing the two-body $\Lambda$N component of the phenomenological Bodmer-Usmani 
potential~\cite{bus88}, and a hyperon-nucleon interaction~\cite{hig14} simulating the scattering phase shifts given 
by NSC97f~\cite{ris99}. The range of applicability of the NSHH method is briefly discussed.
\end{abstract}

\section{Introduction}
In the last decades the physics of hypernuclei has seen increasing interest, testified by the intense experimental activity 
on strange systems. It is then necessary to understand how well theory is able to account for experimental results, 
discriminating among different interaction models. The standard hyperon-nucleon (YN) database comprises 35 selected 
$\Lambda p$ low-energy scattering data and some $\Lambda$N and $\Sigma$N data at higher energies~\cite{kad71} for a 
total of 52 YN scattering data. In comparison the Nijmegen NN scattering database~\cite{sto93} includes over 4300 NN data 
in the range $0\div350$ MeV.

The evidently limited information available for strange nuclear systems highlights the necessity of instruments to test 
the quality of the interaction models, and ab-initio methods are the natural ones. In fact, the accuracy of the results 
can be systematically controlled. This makes the comparison theory-experiment conclusive with respect to the input dynamics. 
In particular, ab-initio methods allow to partially compensate the lack of scattering data by exploiting the experimental 
information on hypernuclear bound states in order to provide new constraints on the YN potential. Therefore, in the strange 
sector, ab-initio calculations for bound states play an even more important role compared to the nuclear case.

The purpose of this contribution is to introduce the NSHH method as a new ab-initio approach in the few-body hypernuclear 
sector. By providing a benchmark study with three other different methods we define its range of applicability. We focus 
on bound-state energies of light hypernuclei from $A=3$ to $A=5$, employing two phenomenological interactions defined in 
configuration space, namely the two-body $\Lambda$N component of the Bodmer-Usmani potential~\cite{bus88}, and the NSC97f
simulated YN potential~\cite{hig14}.

Besides the NSHH, the methods used are the Auxiliary Field Diffusion Monte Carlo method (AFDMC)~\cite{lon13}, 
the Faddeev-Yakubovsky approach (FY)~\cite{yak67} and, by reference, the Gaussian Expansion Method (GEM)~\cite{hig14}. 
Because of the different implementation of such algorithms, we provide a cross benchmark employing the Bodmer-Usmani 
interaction for the first three methods and the NSC97f potential for NSHH, FY and GEM.

\section{The NSHH approach for hypernuclei}
The NSHH method was first introduced by Gattobigio et al.~\cite{gak11} and later extended by Barnea et al.~\cite{deb13} 
in order to treat nuclear systems with realistic NN interactions. In the following we extend NSHH to systems composed 
by two species of fermions, as in the case of hypernuclei.

If $n_1$ is the number of particles of one species and $n_2=A-n_1$ the number of particles of the second species, 
we define the operator $\hat{C}(n_1,n_2)$ as the sum of the Casimir operators $\hat{C}(n_1)$ and $\hat{C}(n_2)$ 
associated, respectively, to the permutation groups $S_{n_1}$ and $S_{n_2}$: 
\begin{equation}
\hat{C}(n_1,n_2)=\hat{C}(n_1)+\hat{C}(n_2)=\sum_{j>i=1}^{n_1} \hat{P}_{ij} + \sum_{j>i=n_1+1}^{A} \hat{P}_{ij}\;,
\end{equation}
where $\hat{P}_{ij}$ is the transposition operator for particles $i$ and $j$. The operator $\hat{C}(n_1,n_2)$ 
commutes with the Hamiltonian of the system and its largest and smallest eigenvalues, $\lambda_s>0$ and $\lambda_a<0$, 
correspond to the symmetric and antisymmetric eigenstates.

The NSHH basis has no permutational symmetry, therefore the action of a permutation operator of two identical particles, 
in general, is not simple. However, by considering the pseudo-Hamiltonian:
\begin{equation}
\tilde H=H+\gamma \hat{C}(n_1,n_2)\;,
\end{equation}
one can choose a suitable value for the $\gamma$ parameter so that the ground-state energy of the physical system becomes 
the lowest eigenvalue of $\tilde H$ subtracted by $\gamma \lambda_a$ (in the case of fermionic systems). The computational 
burden due to the size of the NSHH basis and to the complicated action of the permutation operators can be partly 
overcome by implementing fast procedures to evaluate just the lowest eigenvalues of a matrix (e.g. the Lanczos algorithm). 
In this way the problem of the symmetrization procedure, which is the main source of computational effort in the standard 
HH approach, is removed.

A procedure based on Lee-Suzuki theory~\cite{blo01} is also added to construct an effective two-body interaction in order 
to extend the applicability of the method to a wide range of potentials that present convergence problems when 
treated as bare interactions.

\section{Results}
We studied two different cases, employing two standard versions of the Argonne NN potential in combination with the two 
aforementioned interaction models for the YN force. We computed the total binding energies of nuclei and corresponding 
$\Lambda$-hypernuclei and the $\Lambda$ separation energy $B_\Lambda$, defined as the energy difference between the 
system without and with the $\Lambda$-hyperon.

In the first case the adopted NN potential is the Argonne V4' (AV4')~\cite{wip02}, reprojection of the realistic Argonne 
V18~\cite{wis95} on the first four channels. The electromagnetic part is omitted. The YN interaction is the 
Bodmer-Usmani~\cite{bus88}, which is an Argonne-like potential with two-body $\Lambda$N and three-body $\Lambda$NN components. 
For this study we employed the two-body $\Lambda$N part only. As explained in Ref.~\cite{lon13}, the omission of the 
three-body hyperon-nucleon force in this framework produces overestimated hyperon separation energies, as can also be seen in 
Tab.~\ref{tab-1}. However, the aim of this work is to compare the accuracy of NSHH for hypernuclear systems with other 
ab-initio methods for a given interaction model, rather than reproducing the experimental results. In Tab.~\ref{tab-1} 
we report the complete comparison between NSHH and AFDMC for three- to five-body systems, and an additional comparison 
with Faddeev results in the three-body case~\cite{nogPC}. Green's Function Monte Carlo (GFMC) results for nuclei are also shown~\cite{wip02}.

\begin{table}[htb]
\centering
\caption{Binding and separation energies in MeV for different systems with $A=3-5$. The NN potential is the AV4' 
(no Coulomb force) and the YN potential is the two-body $\Lambda$N Bodmer-Usmani.}
\label{tab-1}
\begin{tabular}{l c c c c c c}
\toprule
\textbf{Interaction} & \textbf{System}   & \textbf{NSHH} & \textbf{AFDMC} & \textbf{FY} & \textbf{GFMC} &\textbf{exp}\\
\midrule                                                                                
AV4'        & $^2_{\phantom{\Lambda}}$H  &               & -2.245(15)     & -2.245(1)   & -2.24(1)      & -2.225 \\
AV4'+Usmani & $^3_\Lambda$H              &  -2.530(3)    & -2.42(6)       & -2.537(1)   &               & \\
            & $B_\Lambda$                &   0.290(3)    &  0.18(6)       &  0.292(1)   &               & 0.13(5) \\
\midrule                                                                                
AV4'        & $^3_{\phantom{\Lambda}}$H  & -8.98(1)      & -8.92(4)       &             & -8.99(1)      & -8.482 \\
AV4'+Usmani & $^4_\Lambda$H              & -12.02(1)     & -11.94(6)      &             &               & \\
            & $B_\Lambda$                &   3.04(1)     &   3.02(7)      &             &               & 2.04(4) \\
\midrule                                                                                
AV4'        & $^4_{\phantom{\Lambda}}$He & -32.89(1)     & -32.84(4)      &             & -32.88(2)     & -28.30 \\
AV4'+Usmani & $^5_\Lambda$He             & -39.54(1)     & -39.51(5)      &             &               & \\
            & $B_\Lambda$                &   6.65(1)     &   6.67(6)      &             &               & 3.12(2) \\
\bottomrule
\end{tabular}
\end{table}

The uncertainties of NSHH results are calculated as the standard deviation on the last four data in the convergence pattern. 
Due to the larger computational effort in reaching full convergence in the four- and five-body case, the error bar is about 
one order of magnitude larger than in the three-body case. 
In particular we have an error estimate of few KeV for both NSHH and FY calculations for $^3_{\Lambda}$H and of 10 keV for 
$^4_{\Lambda}$H and $^5_{\Lambda}$He. More precise calculations with error bars reduced by about a factor ten will be 
calculated in the near future. AFDMC uncertainties are typically larger due to the statistical nature of the method.

The five-body results, both binding and separation energies, are in very good agreement among different methods. For lighter 
systems good agreement is found for $B_\Lambda$, while AFDMC binding energies are $50-100$ keV higher than the corresponding NSHH. 
This is due to technical complications in the AFDMC implementation of the many-body wavefunction for open-shell systems. 
A new way to treat two- and three-body correlations in AFDMC is under study.

Due to the central character of the potentials, the NSHH basis is constrained by the total orbital angular momentum 
$L$ and the spin $S$ of the system, besides the isospin numbers $T$ and $T_z$. The resulting order of magnitude of 
the basis employed to reach the accuracy of the values shown in Tab.~\ref{tab-1} is $10^2$ for the three-body, $10^4$ for the four-body and $10^6$ for the 
five-body case. These dimensions would not be sufficient to reach convergence within a simple variational approach,
due to the strong short-range repulsion of the potential. The Lee-Suzuki procedure generates softer effective interactions, allowing 
for the efficient computation of the NSHH results shown in Tab.~\ref{tab-1}.

In the second case the NN potential is the Argonne V8' (AV8')~\cite{wip02} with no Coulomb force. The employed YN interaction simulates the scattering phase shifts given by NSC97f and it contains a central, a tensor and a spin-orbit term. It includes a $\Lambda-\Sigma$ coupling by taking into account the $\Sigma$ degree of freedom. The AFDMC method has not been extended yet to deal with explicit $\Sigma$. The comparison is then carried out among three methods, namely NSHH, FY~\cite{nogPC} and GEM. Results are shown in Tab.~\ref{tab-2} for three- and four-body systems. The GEM values are taken from Ref.~\cite{hig14}.

\begin{table}[htb]
\centering
\caption{Binding and separation energies in MeV for different systems with $A=3-4$. The NN potential is the Argonne V8' (no Coulomb force). The YN potential is the NSC97f.}
\label{tab-2}
\begin{tabular}{l c c c c c c}
\toprule
\textbf{Interaction} & \textbf{System}  & \textbf{NSHH} & \textbf{FY} & \textbf{GEM} & \textbf{GFMC} & \textbf{exp}\\
\midrule                                                                             
AV8'        & $^2_{\phantom{\Lambda}}$H &               & -2.226(1)   &              &               & -2.225 \\
AV8'+NSC97f & $^3_\Lambda$H             & -2.41(2)      & -2.415(1)   &              &               & \\
            & $B_\Lambda$               &  0.17(2)      & 0.189(1)    & 0.19(1)      &               & 0.13(5) \\
\midrule                                                                             
AV8'        & $^3_{\phantom{\Lambda}}$H & -7.76(0)      &             &              & -7.76(1)      & -8.43(1) \\
AV8'+NSC97f & $^4_\Lambda$H             & -10.05(7)     &             &              &               & \\
            & $B_\Lambda$               &  2.29(7)      &             & 2.33(1)      &               & 2.04(4) \\
\bottomrule
\end{tabular}
\end{table}

The agreement is good both in the three- and four-body case. The NSHH error bars are larger due to approximations made in order to treat the $\Lambda-\Sigma$ mass difference in the definition of the mass-weighted coordinates for the internal motion. An extension of the NSHH method in order to avoid these approximations is under development.

Since both the NN and YN potentials are not central and the $\Sigma$ is treated as explicit degree of freedom, the NSHH Hilbert space is much bigger compared to the previous test based on the Bodmer-Usmani interaction. The basis dimension is one order of magnitude larger and therefore the convergence in this case requires additional computational effort.

\section{Conclusions}
The accuracy of the NSHH approach for three-body systems is good in comparison to other ab-initio methods such as AFDMC, FY and GEM. Its applicability goes beyond $A=3$ systems and it has been tested for $A=4$ and $A=5$. The potentiality of the method is expected to be completely exploited by combining an efficient parallelization procedure in order to deal with larger basis dimensions. This has recently been achieved and the future aim is the study of systems with $5\leq A\leq7$, including cases with strangeness $S=-2$. The present benchmark calculation is intended to be the starting point for the application of the NSHH method to the study of hypernuclear systems.




\end{document}